\documentstyle[a4wide,12pt]{article}
\title{Quantum Algorithm for Hilbert's Tenth Problem}
\author{Tien D Kieu~\footnote{kieu@swin.edu.au}\\
Centre for Atom Optics and Ultrafast Spectroscopy,\\Swinburne
University of Technology, Hawthorn 3122, Australia}
\begin{document}
\maketitle
\begin{abstract}
We explore in the framework of Quantum
Computation the notion of {\em Computability}, which holds a central position in
Mathematics and Theoretical Computer Science.  A quantum algorithm
for Hilbert's tenth problem,
which is equivalent to the Turing halting problem and is known to be mathematically
noncomputable, is proposed where quantum
continuous variables and quantum adiabatic evolution are employed.
If this algorithm could be physically implemented, as much as it is
valid in principle---that is, if certain hamiltonian and its ground state can be 
physically constructed according to the proposal---quantum computability would surpass
classical computability as delimited by the Church-Turing thesis.  It is 
thus argued that computability, and with it the limits of Mathematics,
ought to be determined not solely by Mathematics itself but also by Physical
Principles. 
\end{abstract}
\section{Introduction}
Computation based on the principles of Quantum Mechanics~\cite{qc} has been shown
to offer better performances over classical computation, 
ranging from the square-root improvement in an unstructured search~\cite{grover}
to the exponential gain in the factorisation of integers~\cite{shor}.
However superior in reducing the complexity of hard computation, these quantum 
algorithms and all the others discovered so far are only applicable to the 
classically computable functions.  That leaves untouched the class of 
{\it classically noncomputable} functions, such as the halting problem for Turing 
machines~\cite{recursive}.  It is in fact widely believed that quantum computation 
cannot offer anything new about computability~\cite{QTM}.  Contrary to this, 
we propose that quantum computation may be able to compute the noncomputables, provided 
certain hamiltonian and its ground state can be physically constructed.  
We propose a quantum algorithm for the classically noncomputable
Hilbert's tenth problem~\cite{hilbert} which ultimately links to 
the halting problem for Turing machines in the computation of partial recursive 
functions.

The practical details of implementation the quantum algorithm for this class of problems 
are not considered in this conceptual study, and will be investigated elsewhere.

\section{Hilbert's tenth problem}
At the turn of the last century, David Hilbert listed 23 important problems, among
which the problem number ten could be rephrased as:
\begin{quote}
\it
Given any polynomial equation with any number of unknowns and with integer
coefficients:  To devise a universal process according to which it can be
determined by a finite number of operations whether the equation has
integer solutions.
\end{quote}

This decision problem for such polynomial equations,
which are also known as Diophantine equations,
has eventually been shown in 1970 by Matiyasevich to be 
undecidable~\cite{hilbert, davis} in the Turing sense.  It is consequently
noncomputable/undecidable in the most general sense if one accepts, as 
almost everyone does, the Church-Turing thesis of computability.
Since exponential Diophantine, with the unknowns in the exponents
as in the example of the Fermat's last theorem, can be shown to be
Diophantine with supplementary equations, the study of Diophantine equations 
essentially covers the class of partial recursive functions, which is at 
the foundations of classical algorithms.  
The undecidability result is thus singularly important:  Hilbert's tenth
problem could be solved if and only if the Turing halting problem could be.

\section{Turing halting problem}
The halting problem for Turing machines is also a manifestation of undecidability:  
a Turing computation is equivalent to the computation of a partial recursive function, 
which is only defined for a subset of the integers; as this domain is
classically undecidable, one cannot always tell in advance whether the Turing machine 
will halt (that is, whether the input is in the domain of the partial recursive function) 
or not (when the input is not in the domain).

A version of the proof of the unsolvability of the halting problem
based on the Cantor diagonal argument goes as follows.  The proof is by contradiction 
with the assumption of the existence of a computable halting function $h(p,i)$ which has 
two integer arguments - $p$ is the G\"odel encoded integer number for the 
algorithm and $i$ is its (encoded) integer input:
\begin{eqnarray}
h(p,i) &=& \left\{\begin{array}{ll}
0 & \mbox{if $p$ halts on input $i$}\\
1 & \mbox{if $p$ does not}
\end{array} \right.  \label{h}
\end{eqnarray}
One can then construct a program $r(n)$ having one integer argument $n$ in
such a way that it calls the function $h(n,n)$ and 
\begin{eqnarray}
\left\{
\begin{array}{l}
\mbox{$r(n)$ halts if $h(n,n) = 1$} \nonumber\\
\mbox{$r(n)$ loops infinitely (i.e., never stops) otherwise.}
\nonumber
\end{array}
\right.
\end{eqnarray}
The application of the halting function $h$ on the program $r$ and input $n$ 
results in
\begin{eqnarray}
h(r,n) &=& \left\{\begin{array}{ll} 
0 & \mbox{if $h(n,n)=1$}\\
1 & \mbox{if $h(n,n)=0$}
\end{array}\right. \label{contradiction}
\end{eqnarray}
A contradiction is clearly manifest once we put $n=r$ in the last equation
above.  

The construction of such program $r$ is transparently
possible, unless the existence of a computable $h$ is wrongly assumed.  
Thus the contradiction discounts the assumption that there is a classically 
algorithmic way to determine whether any arbitrarily given program with 
arbitrary input will halt or not.

This contradiction argument might be side stepped if we distinguish and
separate the two classes of quantum and classical algorithms.  A {\it
quantum} function $qh(p,i)$, similar to eq.~(\ref{h}), can conceivably 
exist to determine whether any classical program $p$ will halt on any 
classical input $i$ or not.  The contradiction in eq.~(\ref{contradiction}) 
would be avoided if the quantum halting $qh$ cannot take as argument the 
modified program $r$, which is now of {\it quantum} character because it now
has quantum $qh$ as a subroutine.  This will be the case if $qh$
can only accept integer while quantum algorithms, 
with proper definitions, cannot in general be themselves encoded as integers.  
It is clear even in the case of a single qubit
that the state $\alpha|0\rangle + \beta|1\rangle$ cannot be encoded as integers 
for all $\alpha$ and $\beta$ - simply because of different cardinalities.  
In fact, the no-cloning theorem~\cite{no-clone} of quantum mechanics does 
restrict the type of operations available to quantum algorithms.

In essence, the way we will break the self-referential reasoning here by the
differentiation between quantum and classical algorithms is similar to
the way John von Neumann and Bertrand Russell resolved the set theory paradox (to do with
``The set of all sets which are not members of themselves") by the
introduction of classes as distinct from sets.  (For other lines of
arguments, see~\cite{ordkieu}.)

\section{An observation}
It suffices to consider nonnegative solutions, if any, of a
Diophantine equation.  
Let us consider the example
\begin{eqnarray}
(x+1)^3 + (y+1)^3 - (z+1)^3 + cxyz = 0, && c\in Z,
\label{eq}
\end{eqnarray}
with unknowns $x$, $y$, and $z$.
%which is a particular case of the Fermat's last theorem.  
To find out
whether this equation has any nonnegative integer solution by quantum
algorithms, it requires the realisation of a Fock space built out of the
``vacuum" $|0_a\rangle$ by repeating applications of the creation operators
$a^\dagger_x$, $a^\dagger_y$ and $a^\dagger_z$, similarly to that of the
3-D simple harmonic oscillators.
\begin{eqnarray}
[a_j, a^\dagger_j] = 1 &&\mbox{for $j = x, y, z$},\\ \nonumber
[a_k, a_j] = [a_k, a^\dagger_j] = 0 &&\mbox{for $j \not = k$}.
\end{eqnarray}
Upon this Hilbert space, we construct the hamiltonian corresponding
to~(\ref{eq})
\begin{eqnarray}
H_P &=& \left((a^\dagger_x a_x+1)^3 + (a^\dagger_y a_y+1)^3 - (a^\dagger_z
a_z+1)^3 + c(a^\dagger_x a_x)(a^\dagger_y a_y)(a^\dagger_z a_z) \right)^2,
\nonumber
\end{eqnarray}
which has a spectrum bounded from below -- semidefinite, in fact.  

Note that the operators $N_j = a^\dagger_j a_j$ have only 
nonnegative integer eigenvalues $n_j$, and that $[N_j, H_P] = 0 = 
[N_i, N_j]$ so these 
observables are compatible -- they are simultaneously measurable.   The 
ground state $|g\rangle$ 
of the hamiltonian so constructed has the properties
\begin{eqnarray}
N_j|g\rangle &=& n_j|g\rangle, \nonumber\\
H_P|g\rangle &=& \left((n_x+1)^3 + (n_y+1)^3 - (n_z+1)^3 + cn_xn_yn_z
\right)^2|g\rangle \equiv E_g |g\rangle,
\nonumber
\label{eigenvalues}
\end{eqnarray}
for some $(n_x,n_y,n_z)$.

Thus a projective measurement of the energy $E_g$  of the ground state 
$|g\rangle$ will yield 
the answer for the decision problem: The Diophantine equation has at least one
integer solution if and only if $E_g = 0$, and has not otherwise.  (If
$c=0$ in our example, we know that $E_g > 0$ from Fermat's last theorem.)

If there is one unique solution then the projective
measurements of the observables corresponding to the operators
$N_j$ will reveal the values of various unknowns.  If there
are many solutions, finitely or infinitely as in the case of $x^2 + y^2 -
z^2 = 0$, the ground state $|g\rangle$ will be a linear superposition of 
states of the form $|n_x\rangle\otimes|n_y\rangle\otimes|n_z\rangle$, where
$(n_x,n_y,n_z)$ are the solutions. In such situation, the measurement may 
not yield all the solutions.  However, finding all the solutions is not
the aim of a decision procedure for this kind of problem.  

Notwithstanding this, measurements of $N_j$ of the ground state would always 
yield some values $(n_x,n_y,n_z)$ and a straightforward substitution would 
confirm if the equation has a solution or not.  Thus the measurement on the 
ground state either of the energy, provided the zero point can be calibrated,
or of the number operators will be sufficient 
to give the result for the decision problem.  

The quantum algorithm with the ground-state oracle is thus clear:
\begin{enumerate}
\item  Given a Diophantine equation with $K$ unknowns $x$'s
\begin{eqnarray}
D(x_1,\cdots,x_K) &=& 0,
\end{eqnarray}
we need to simulate on some appropriate Fock space
the quantum hamiltonian 
\begin{eqnarray}
H_P &=& \left(D(a^\dagger_1 a_1,\cdots, a^\dagger_K a_K) \right)^2.
\end{eqnarray}
\item  If the ground state could be obtained with high probability,
measurements of appropriate observables would provide the answer for
our decision problem.
\end{enumerate}
The key ingredients are the availability of
a countably infinite number of Fock states, the ability to construct/simulate
a suitable hamiltonian and to obtain/identify its ground state via quantum measurements.
As a counterpart of the semi-infinite tape of a Turing machine, the Fock 
space is employed here instead of the qubits of the more well-known
model of quantum computation.  Its advantage over the infinitely many 
qubits which would otherwise be required is obvious.  

\section{Some preliminary comments} 
We do not look for the zeroes of the polynomial,
$D(x_1,\cdots,x_K)$, which may not exist, but instead search for the absolute 
minimum of its square which exists,
\[ 0\le \min \left(D(x_1,\cdots,x_K)\right)^2 \le \left(D(0,\cdots,0)\right)^2, \]
and is finite because $\lim_{x\to\infty} \left(D(x_1,\cdots,x_K)\right)^2$
diverges.

While it is equally hard to
find either the zeroes or the absolute minimum in classical computation, 
we have converted the problem to the realisation of the ground state of
a quantum hamiltonian and there is no known quantum principle against
such act.  In fact, there is no known physical principles 
against it.  Let us consider the three laws of thermodynamics concerning 
energy conservation, entropy of closed systems and the unattainability
of absolute zero temperature.  The energy involved in our algorithm
is finite, being the ground state energy of some hamiltonian.  The
entropy increase which ultimately connects to decoherence effects is a
technical problem for all quantum computation in general, and we will
discuss this further below.  As we will never obtain the absolute zero
temperature, we only need to satisfy ourselves that the required ground
state can be achieved with a more-than-even chance.  Then there is a
probability boosting technique, see later, to bring that chance to as 
closed to unity as one pleases.

It may appear that even the quantum process can only explore a
finite domain in a finite time and is thus no better than a classical
machine in terms of computability.  But there is a crucial
difference.

In a classical search even if the global minimum is come across, it cannot 
generally be 
proved that it is the global minimum (unless it is a zero of the Diophantine 
equation).  Armed only with mathematical logic, we would still have to 
compare it with all other numbers from the 
infinite domain yet to come, but we obviously can never complete this 
comparison in finite time --thus, mathematical noncomputability.

In the quantum case, the global minimum is encoded in the ground
state.  Then, by energy tagging, the global minimum can be found in finite 
time and confirmed, if it is the ground state that is obtained at the 
end of the computation.  And the ground state may be identified and/or verified by 
physical principles.  These principles are over and above 
the mathematics which govern the logic of a classical machine and help
differentiating the quantum from the classical.  Quantum mechanics could 
``explore" an infinite domain, but only in the sense that it can select, 
among an infinite number of states, one single state (or a subspace in case 
of degeneracy) to be identified as the ground state of some given hamiltonian
(which is bounded from below).  
This ``sorting" can be done because of energetic reason, which is a physical 
principle and is not available to mathematical computability.  

%\section*{General quantum computation}
On the other hand, our proposal is apparently in
contrast to the claim in~\cite{QTM} that quantum Turing machines
compute exactly the same class of functions as do Turing machines, albeit 
perhaps more efficiently.  We could only offer here some speculations about
this apparent discrepancy.  The quantum Turing machine approach is a
direct generalisation of that of the classical Turing machines but with qubits
and some universal set of one-qubit and two-qubit unitary gates to build up,
step by step, dimensionally larger, but still dimensionally finite unitary operations.  
This universal set is chosen on its
ability to evaluate any desirable classical logic function.
Our approach, on the other hand, is from the start 
based on infinite-dimension hamiltonians acting on some Fock space
and also based on the special properties and unique status of their ground states.  
The unitary 
operations are then followed as the Schr\"odinger time evolutions.  Even at the 
hamiltonian level higher orders of the operators $a$ and $a^\dagger$, i.e. 
not just two-body but many-body interactions in a sense, are already present.  
This proliferation, which is even more pronounced at the level of the time-evolution 
operators, together with the infinite dimensionality and the unique 
energetic status of the vacuum could be the reasons behind the 
ability to compute, in a finite number of steps, what the dimensionally
finite unitary operators of 
the standard quantum Turing computation cannot do in a finite number of steps. 
Note that it was the general hamiltonian computation that was discussed 
by Benioff and Feynman~\cite{early, early2} in the conception days of quantum computation.

Indeed, Nielsen~\cite{nielsen} has also found no logical contradiction
in applying the most general quantum mechanical principles to the
computation of the classical noncomputable, unless certain Hermitean
operators cannot somehow be realised as observables or certain unitary
processes cannot somehow be admitted as quantum dynamics.  And up to now
we do not have any evidence nor any principles that prohibit these kinds
of observables and dynamics.  (Ozawa~\cite{ozawa} has produced some counter arguments but we think they are not
quite applicable here.  See~\cite{ordkieu}.)

Our general algorithm above could be realised by, but in no way restricted to,
the following methods to simulate the required hamiltonian and to
obtain the ground state adiabatically.

\section{Simulating the hamiltonians}
One way to construct any suitable hamiltonian so desired is through
the technique of ref.~\cite{continuous}.  We consider the hermitean 
operators, where $j$ 
is the index of the unknowns of the Diophantine equation,
\begin{eqnarray}
X_j &=& \frac{1}{\sqrt 2}(a_j + a^\dagger_j), \nonumber \\
P_j &=& \frac{i}{\sqrt 2}(a_j - a^\dagger_j),\\   \nonumber 
 [P_j, X_k] &=& i\delta_{jk}.
\end{eqnarray}
Together with the availability of the fundamental hamiltonians 
\begin{eqnarray}
X_j, P_j, (X^2_j + P^2_j), \pm(X_kP_j + P_jX_k), \mbox{ and } (X^2_j + P^2_j)^2
\label{basic}
\end{eqnarray}
one could construct the unitary time evolutions corresponding to hamiltonians
of arbitrary hermitean polynomials in $\{X_j,P_j\}$, and hence in $\{a^\dagger_j
a_j\}$, to an arbitrary degree of accuracy.  
These fundamental hamiltonians correspond to translations,
phase shifts, squeezers, beam splitters and Kerr nonlinearity.

With the polynomial hamiltonian constructed, we need to obtain its ground 
state.  Any approach that allows us to access the ground state will
suffice.  One way is perhaps to use that of quantum annealing or 
cooling~\cite{qannealing}.  Another way
is to employ the quantum computation method of adiabatic 
evolution~\cite{adiabatic}.

\section{Adiabatic quantum evolution}
In the adiabatic approach, one
starts with a hamiltonian $H_I$ whose ground state $|g_I\rangle$ is 
readily achievable.  Then one forms the slowly varying hamiltonian
${\cal H}(s)$, $s=\frac{t}{T}$, which interpolates between $H_I$ 
and $H_P$ in the time interval $t\in[0,T]$
\begin{eqnarray}
{\cal H}(s) &=& \left(1-s\right)H_I + 
s H_P.
\label{hamiltonian}
\end{eqnarray}
Note that we can replace this linear interpolation by some non-linear
one provided the conditions of the adiabatic theorem are observed.
According to this theorem, the initial ground state will evolve into our
desirable ground state $|g\rangle$ up to a phase:
\begin{eqnarray}
\lim_{T\to\infty}{\cal T}\exp\left\{ -iT\int_0^1 {\cal H}(\tau) 
d\tau\right\}|g_I\rangle &=& {\rm e}^{i\phi}|g\rangle.
\label{7}
\end{eqnarray}
For the hamiltonian~(\ref{hamiltonian}), an estimate of the time
$T$ after which
the system remains with high probability in the ground state is
\begin{eqnarray}
T &\gg& \frac{\parallel H_I - H_P \parallel}{g^2},
\label{adiab}
\end{eqnarray}
with
\begin{eqnarray}
\parallel H_I - H_P \parallel &\equiv& \max_{0\le t \le T} \left|\langle 
e(t)|(H_I - H_P)| g(t)\rangle\right|,
\label{norm}
\end{eqnarray}
and
\begin{eqnarray}
g &\equiv& \min_{0\le t \le T} \left(E_e(t) - E_g(t)\right),
\label{gap}
\end{eqnarray}
where $|g(t)\rangle$ and $|e(t)\rangle$ are respectively the instantaneous
ground state and the first excited state of~(\ref{hamiltonian}) with
instantaneous eigenvalues $E_g(t)$, $E_e(t)$.

The time-ordering operator on the left hand side of~(\ref{7}) can be 
approximated as
\[\exp\{-iT{\cal H}(\tau_N)\Delta\tau\}\cdots\exp\{-iT{\cal H}(\tau_1)
\Delta\tau\}\]
for~\cite{adiabatic}
\begin{eqnarray}
N\Delta\tau &=& 1,\nonumber\\
\Delta\tau\parallel H_I - H_P \parallel &\ll& 1.
\end{eqnarray}
Note that we have employed here the ``norm'' $\parallel.\parallel$ as 
defined in~(\ref{norm}) for the various hamiltonians which are unbounded 
from above.  This norm is the relevant measure for the problem is only concerned
with the lowest states of the interpolating hamiltonian~(\ref{hamiltonian}).
In each interval $\Delta\tau$, the unitary
operators $\exp\{-i{\cal H}(\tau_k)T\Delta\tau\}$, for $k = 1,\cdots, N$,
can be expressed through the subdivision of $T\Delta\tau$ into $m$ subintervals 
of sufficiently small size $\delta s$ satisfying $m\delta s = T\Delta\tau$,
\[\exp\{-i{\cal H}(\tau_k)T\Delta\tau\} = \left(\exp\{-i{\cal H}(\tau_k)\delta s\}
\right)^m.\]
Each of the $m$ factors on the right hand side of the last expression can be now
simulated through the approach of~\cite{continuous}, where it was shown that the 
number of steps $M$ grows as a small polynomial in the order of the 
polynomial in the hamiltonian ${\cal H}(\tau_k)$ to be simulated, the accuracy 
to be enacted, and the time interval $T\Delta\tau$ over which it is to be applied.  

In this way, the requirements of the adiabatic conditions on the one hand and 
of, on the other hand, the simulations of the hamiltonians in the time interval 
$T\Delta\tau$ can be satisfied.

\section{An adiabatic algorithm}
In order to solve Hilbert's tenth problem we need on the one hand such time-dependent physical 
(adiabatic) processes.  On the other hand, the theory of Quantum
Mechanics can be used to identify the ground state through the usual
statistical predictions from the Schr\"odinger equation with a finitely truncated number of energy 
states of the time-dependent Hamiltonian ${\cal H}(t/T)$.  This way, we can overcome the 
problem of which states are to be included in the truncated basis for a numerical study 
of Quantum Mechanics.  This also reconciles with the Cantor diagonal arguments which state that 
the problem could not be solved entirely in the framework of classical computation. 

Below is an algorithm~\cite{Kieu} based on this philosophy of exploiting the interplay between the
presumably infinite physical world and the theory of Quantum Mechanics
calculated in a finite manner on Turing machines.  The algorithm
presented may not be the most efficient; there could be many other
variations making better use of the same philosophy.

It is in general easier to implement some Hamiltonian
than to obtain its ground state.  We thus should start the
computation in yet a different and readily obtainable initial
ground state, $|g_I\rangle$, of some initial 
Hamiltonian, $H_I$, then deform this Hamiltonian in a time $T$
into the Hamiltonian whose ground state is the desired one, through a 
time-dependent process represented by the interpolating Hamiltonian ${\cal H}(t/T)$.

In this approach, inspired by the quantum adiabatic approach, one
starts, for example, with a Hamiltonian $H_I$,
\begin{eqnarray}
H_I = \sum_{i=1}^K (a^\dagger_i -\alpha^*_i)(a_i - \alpha_i),
\label{init}
\end{eqnarray}
which admits the readily achievable coherent state 
$|g_I\rangle = |\alpha_1\cdots\alpha_K\rangle$ as the ground state.  
Then, one forms the time-dependent Hamiltonian
${\cal H}(t/T)$ in~(\ref{hamiltonian}), which interpolates in the time interval 
$t\in[0,T]$ between the initial $H_I$ and $H_P$.
 
\begin{itemize}
\item {\em Step 0:}  Choose an evolution time $T$, a probability $p$ which can be 
made arbitrarily closed to unity, and an accuracy $0<\epsilon<1$ which can be made 
arbitrarily small.
\item  {\em Step 1 (on the physical apparatus):}  Perform the {\it physical} 
quantum time-dependent process which is governed by the time-dependent Hamiltonian
${\cal H}(t/T)$ and terminates after a time $T$.   Then, by projective
measurement (either of the observable $H_p$ or the number operators $\{N_1, \ldots, N_K\}$) 
we obtain some state of the form $|\cdots n_i \cdots\rangle$, $i=1,\cdots,K$.
\item {\em Step 2 (on the physical apparatus):}  Repeat the physical process in {\em Step 
1} a number of times, $L(\epsilon,p)$, to build up a histogram of measurement frequencies 
(for all the states obtained by
measurement) until we get a probability distribution $P(T;\epsilon)$ at
the time $T$ with an accuracy $\epsilon$ for all the measured states.
The convergence of this repetition process is ensured by the Weak Law of Large Numbers in
probability theory~\cite{law}.  (An overestimate of the number of repetitions is $L\ge 1/(\epsilon^2(1-p))$.)
Note the lowest energy state so obtained, $|\vec{n}_c\rangle$, as the candidate ground state.

\item  {\em Step 3 (on the classical computer):}  Choose a
truncated basis of $M$ vectors made up of $|\alpha_1\cdots\alpha_K\rangle$ and its
excited states by successive applications of the displaced creation
operators $b^\dagger_i \equiv (a^\dagger_i -\alpha^*_i)$ on the initial state. 
\item {\em Step 4 (on the classical computer):}  Solve the 
Schr\"odinger equation in this basis for $\psi(T)$, with the initial state
$\psi(0) = |\alpha_1\cdots\alpha_K\rangle$, to derive a probability
distribution $P_{\rm est}(T;M)$ (through $|\langle\psi(T)|\cdots n_i
\cdots\rangle|^2$)
which is similar to that of {\em Step 2} and which
depends on the total number $M$ of vectors in the truncated basis.
\item {\em Step 5 (on the classical computer):}  If the two probability distributions are not uniformly
within the desired accuracy, that is, $|P_{\rm est}(T;M) -
P(T;\epsilon)|>\epsilon$, we enlarge the truncated basis by increasing the size $M$ 
and go back to the {\em Step 4} above.
\item  {\em Step 6 (on the classical computer):}  If the two probability distributions are uniformly
within the desired accuracy, that is, $|P_{\rm est}(T;M) -
P(T;\epsilon)|<\epsilon$, then use this truncated basis to diagonalise $H_P$ to yield, within an 
accuracy which can be determined from $\epsilon$, the approximated ground state $|g'\rangle$ and its
energy $E_{g'}$.
\item {\em Step 7 (on the classical computer):} 
We can now estimate in this truncated basis the gap between the ground state and the first excited
state.  From this gap, we can make use of the quantum adiabatic 
theorem and choose a time $T$ such that 
the system has a high probability to be in the ground state
\[ \left||\langle g'|\psi(T)\rangle|^2-1\right|<\epsilon. \]
We then go back to {\em Step 1} with this choice of $T$, which is
to amplify and thus confirm the candidate ground state as the real
ground state.
\end{itemize}
Our point is on computability
and not on computational complexity, which depends on individual
polynomials.  Computability is based on the arguments that the
adiabatic time $T$ is {\em finite}~\cite{ruskai, kieu2} (for a high probability of achieving the
ground state) and that the ground state can be {\em verified} by employing the
theory of Quantum Mechanics.  As long as the energy gap is
finite so is the computational time.  In contrast,
the most general classical algorithm for Hilbert's tenth problem (by systematic substituting
in integers of larger and larger magnitudes) cannot solve it in principle even
allowing for exponentially grown, but finite, amount of time -- unless
{\em infinite} amount of time were available, which it is not.

Given a Diophantine equation, the  substitution
in integers of larger and larger magnitudes is not satisfactory as we do not know when
the substitution should be terminated.  Likewise, if we want to numerically simulate the
quantum algorithm proposed, we would have to use a finitely truncated having $M$
vectors.  But we face the 
same problem of not knowing which $M$ to choose in general.  That is why 
the problem is noncomputable on classical computers.

Even we can estimate $T$, as in the appendix, for some starting point of the
adiabatic computation as we might not be able in general to exactly know $T$ {\em a priori}.  
Were the required adiabatic time $T$ somehow known exactly within classical computation 
and without the help of quantum computation
then the problem might be solved classically.  
But as Hilbert's tenth problem cannot be
solved by classical computation, we will have to resort to quantum
computation without {\em a priori}
knowing exactly the time $T$, except the knowledge that it is {\em finite}.
Constructive
logicians~\cite{calude} allow for this algorithmic situation under the 
so-called Markov's Principle.

\section{Discussion of the algorithm}
The quantum algorithm above
can be proved to terminate (even though it could be after a very long time) 
and give us the decision result for Hilbert's tenth problem.

The real spectrum of $H_p$ is of integer
values (in suitable units), and that is what we also get from
measurement.  But the spectrum calculated from a finitely truncated
basis is not of integer values and will fluctuate with fluctuation size
depends on the size of the truncated basis employed.  The accuracy size 
$\epsilon$ of the measured probability distribution is chosen such that the off-set 
of the ground state energy $\delta$ should allow us to conclude whether the ground state 
energy $E_{g'}$ is zero or not.  ($\delta$ is in general
a function of $\epsilon$ and $T$.)
\begin{eqnarray}
E_{g'} &=& \langle g'|H_P|g'\rangle,\nonumber\\
&=& E_c + \langle r|H_P|r\rangle + 2{\rm Re}\langle r|H_P|n_c\rangle,\nonumber\\
&=& E_c + \langle r|H_P|r\rangle + 2E_c({\rm Re}\langle r|n_c\rangle),\nonumber\\
&\equiv& E_c + \delta(\epsilon),
\end{eqnarray}
where $|r\rangle \equiv |g'\rangle - |n_c\rangle$ and $H_P|n_c\rangle = E_c |n_c\rangle$. 

The termination of our algorithm is obtained if and when the adding of higher 
$b$-number states (those created from the coherent state by the
application of the creation operator $b^\dagger_i \equiv (a^\dagger_i -
\alpha^*_i)$) to the truncated basis does not change the approximated ground state of $H_P$
beyond certain range of accuracy,
\begin{eqnarray}
|\delta(\epsilon)| &\le& E_c \not = 0.
\label{condition}
\end{eqnarray}

Even we can prove that the approximated ground state $|g'\rangle$ and its energy will 
eventually converge to its true values, the mathematical noncomputability 
results from the fact that their rates of convergence are unknown.  
Thus, we might not be able to use mathematical reasoning alone
to determine when to stop adding more states to the truncated basis in
order to approximate the ground state correctly.  Different truncated
bases would give {\em some} estimates for the ground state but we have 
no control over these estimates and no idea how good they are.  
They could be anywhere in relation to the true values.  This is nothing but 
mathematical noncomputability.  (However, this quantum algorithm has inspired us
to reformulate Hilbert's tenth problem with continuous variables~\cite{kieu2}, 
and the mathematical computibility of this reformulation, or lack of it, should be 
investigated further.)

To know when the truncated basis is sufficiently
large to have the estimated ground state values within any given
accuracy, that is, to regain computability, we have to exploit the
measurability of physical processes.  Because of this measurability we
can estimate the true accuracy of our measured values.   Then a
comparison of results from the Schr\"odinger equation to these
measurable quantities will help determining the accuracy of results from
the equation, that is, regaining the lost computability through the
physical world, presumably infinite.

% We have
% \begin{eqnarray}
% |\delta(\epsilon)| &\le& |\langle r|H_P|r\rangle| + 2E_c|{\rm Re}\langle r|n_c\rangle|,\nonumber\\
% &\le& \left\| |r\rangle\right\|.\left\| H_P|r\rangle\right\| + 2E_c\left\| |r\rangle\right\|, \nonumber\\
% &\le& \left\| |r\rangle\right\|\left(\left\| H_P|r\rangle\right\| + 2E_c \right).
% \end{eqnarray}
% With the decomposition
% \begin{eqnarray}
% |g'\rangle &=& \sum d_i |E_i\rangle,
% \end{eqnarray}
% we can have the estimation
% \begin{eqnarray}
% \| H_P|r\rangle\| &=& \left\|\sum_{i\not= c} d_i E_i |E_i\rangle + E_c (d_c -1)|n_c\rangle\right\|,\nonumber\\
% &\le& \sqrt{\langle g' |E^2|g'\rangle} +2 E_c,
% \end{eqnarray}
% From which
% \begin{eqnarray}
% |\delta(\epsilon)| &\le& \|| |r\rangle\||\left( 4E_c + \sqrt{\langle g'
% |E^2|g'\rangle} \right).
% \end{eqnarray}
% The right hand side of this last inequality is a function of $\epsilon$
% and some $\epsilon$ can be {\em empirically} chosen such that either $E_c=0$ 
% (upon which we can conclude that the Diophantine equation under consideration has
% some integer solution)
% or the inequality~(\ref{condition}) is satisfied (upon which we can draw
% the conclusion that the Diophantine equation under consideration has no
% integer solution).

\section{Decoherence and error correction}
Our approach above is in fact a combination of the quantum computation of
continuous variables and of adiabatic evolution.  There exists some
error correction protocol for continuous variables~\cite{error} which
could be of help here to protect the wave functions from decoherence.  
However, the adiabatic computation we exploit is quite robust in 
general~\cite{robust}.  An imperfect conventional quantum algorithm
might have different sorts of errors than an imperfect adiabatic process,
where the system is kept close to the instantaneous ground state over 
time.  Decoherence by the environment inducing
transition between levels could be controlled in principle at a temperature
that is small compared to the gap~(\ref{gap}), given its estimate.  Errors
introduced by the hamiltonian simulation may result in a hamiltonian different
from~(\ref{hamiltonian}) but in a form ${\cal H}(t/T) + K(t)$.  Recent numerical
study~\cite{robust} of small systems has in fact indicated that the adiabatic
computation is interestingly robust even for fairly large
$K(t)$, provided $K(t)$ varies either sufficiently slowly or sufficiently rapidly 
(which is more likely to be the case considered here due to the nature of our 
hamiltonian simulations).  

\section{Probability boosting}
Due to the particular nature of the present scheme, the various 
approximations result in some probability that the final measurement 
will not find the system in the desired ground state; but appropriate 
choices of the various time parameters could increase 
the success probability of the algorithm to more than even.
We spell out explicitly here the probability boosting technique, as
mentioned in~\cite{QTM}, that one could subsequently apply.  

In our computation, adiabatic or otherwise,
the end result, starting from some initial state $|g_I\rangle$, 
may be contaminated with some excited states other than
the desirable ground state $|g\rangle$,
\begin{eqnarray}
|g_I\rangle &\stackrel{\rm time}{\longmapsto}& a |g\rangle + b |e\rangle.
\label{contam}
\end{eqnarray}
If $|a| > |b|$, that is, if there is a better-than-even chance to obtain
$|g\rangle$ by measurement then one can boost the success probability to
arbitrarily closed to unity by performing a concatenated computation
over $l$ Hilbert spaces
\begin{eqnarray}
|g_I\rangle_1\otimes\cdots|g_I\rangle_l &\stackrel{\rm time}{\longmapsto}& 
{\cal N}\sum_{p=0}^l a^{p}b^{l-p} |e\rangle_1\otimes\cdots
\otimes|g\rangle_{k_1}\otimes\cdots\otimes|g\rangle_{k_p}
\otimes\cdots\otimes|e\rangle_{l},
\label{concat}
\end{eqnarray}
where ${\cal N}$ is the normalising factor.  Let us consider the
majority amplitudes, when more than half of the $l$ Hilbert spaces return
the correct results, ${\cal N}a^pb^{l-p}, p > \frac{l}{2}$;
and the minority amplitudes, ${\cal N}a^qb^{l-q}, q < \frac{l}{2}$.
The ratio of the majority over the minority, $(a/b)^{p-q}$, is clearly
boosted  for sufficiently large $l$ and for $p\sim O(l)$ and $q\sim O(1)$.  
The probability distribution for a measurement of the 
end state in~(\ref{concat}), as a consequence,
is exponentially dominated by the majority results.  In
other words, the probability to obtain a majority result which contains
the true ground state can be made arbitrarily closed to unity, 
$(1-\epsilon')$, provided $|a| > |b|$ and $l > -C\log\epsilon'$.  
However, the decoherence control for such
$l$ concatenated Hilbert spaces will be more crucial.

\section{Concluding remarks}
% That is, there might be some
% fundamental physical principles, not those of practicality, which prohibit
% the implementation.  For example, a precision of, say, $10^{-20}$ would be 
% required but that is not allowed {\it in principles}, and not just because 
% of present-day limitation on technology.  Or, there might be not enough 
% physical resources (ultimately limited by the total energy and the lifetime 
% of the universe) to satisfy the execution of the algorithm.  (In this case 
% it is likely that the Turing program in consideration, even if it eventually 
% halts upon some input, would take a running time longer than the lifetime 
% of the universe.)  In either cases, the whole exercise is still very 
% interesting as the unsolvability of those problems and the limit of mathematics
% itself are also dictated by physical principles and/or resources.
In this paper, we consider and emphasise on the 
issue of computability in principle, not that of computational complexity.  
This attempt of broadening of the concept of effective computability, taken 
into account the quantum mechanical principles, has been argued to be able in
principle
to decide the classically/Turing undecidables, Hilbert's 
tenth problem and thus the Turing halting problem in this instance. 
If this is realisable, and we don't have any evidence of fundamental nature
to the contrary, the Church-Turing thesis should be modified accordingly. 

In summary, we have encoded the answer to the question about the existence 
or lack of non-negative integer solutions for any Diophantine equation 
into the of ground state of some relevant Hamiltonian.  
The key factor in the ground state verification is {\em the probability distributions}, 
which not only can be calculated in numerical Quantum Mechanics 
(with a truncated basis) but also are 
measurable in practice.  After all, probability distributions are also
physical observables.
However, in using the probability distributions as the identification criteria, we have 
to assume that Quantum Mechanics is able to describe Nature correctly to the precision 
required.  Note also that we have here a peculiar situation in which the computational 
complexity, that is, the computation time, might not be known exactly {\em before}
carrying out the quantum computation -- although it can be estimated approximately (see the appendix below). 

On the other hand, if for any reasons the algorithm is not implementable 
because of physical principles and/or physical resources 
then it would be an example of information being limited by
physics, rather than by logical arguments alone.

Our study is an illustration of ``Information is physical" 
(see~\cite{calude2} for another quantum mechanical approach 
and ~\cite{relativity} for where the theory of General Relativity 
is also exploited for the computation of Turing noncomputables).  

That some generalisation of the notion of computation could help solving 
the previous undecidability/noncomputability has been recognised %~\cite{peter}
in mathematics and was considered by Kleene as quoted in~\cite{recursive}. 
But this has not been realisable until now simply because of the non-recognition of 
quantum physics as the missing ingredient.  Our quantum algorithm could in fact 
be regarded as an infinite search through the integers in a finite amount of time, 
the type of search required by Kleene to solve the Turing halting problem.

Our decidability study here only deals with the property of being Diophantine,
which does not cover the property of being arithmetic in general, and as
such has no direct consequences on the G\"odel's Incompleteness theorem~\cite{kieu3}.
However, it is conceivable that the G\"odel's theorem may lose its
restrictive power once the concept of proof is suitably generalised with
quantum principles.

\section*{Note added in proof}
The author has recently obtained a criterion based on the final propbability
distribution, which can be measured to arbitrary precision, for identifying
the ground state of $H_P$ at time $T$.  See~\cite{kieunumerical} for more details.

\section*{Acknowledgements}  
I am indebted to Alan Head for discussions, 
comments and suggestions.  I would also like to thank Cristian Calude, John Markham,
Boris Pavlov, Andrew Rawlinson and Khai Vu for discussions; Gregory Chaitin, Bryan Dalton, Ray Sawyer, 
Boris Tsirelson and Ray Volkas for email correspondence; and K. Svozil, G. Etesi 
and I. N\'emeti
for bringing their works to my attention.  % Comments by anonymous referees are also gratefully acknowledged.

\appendix
\section*{Appendix: Gap estimation}
\renewcommand{\theequation}{\roman{equation}}
\setcounter{equation}{0}
The question of computational complexity, 
i.e. how large the adiabatic computational time $T$ is for
a high probability of measurement success, is dependent 
on $H_P$, i.e. on the specific Diophantine equation in question, and on the 
initial hamiltonian $H_I$.  Some estimate for the energy gap~(\ref{gap}) is 
desirable, but not necessary as discussed above.  In this
appendix we propose such an estimate for the {\em Step 0} of the quantum
algorithm.

It has been shown elsewhere~\cite{ruskai, kieu2} that in general
there should be no level crossing for the ground state
except at the end points $t = 0, T$ where the adiabatic process can start or end with 
some obvious symmetry.
Furthermore, the freedom in choosing
the initial hamiltonians $H_I$ and their ground states and in
performing different adiabatic interpolations, not just as linear as 
in~(\ref{hamiltonian}), might be exploited to enable the gap enlargement and to
speed up the computation.

We shall employ the 
simple harmonic oscillators, i.e. gaussian approximations,
to obtain an estimate for the energy gap.

For the various $a_i$, $a^\dagger_i$ appearing in the interpolating 
hamiltonian ${\cal H}(s)$~(\ref{hamiltonian}) at the time instant $s$,
we use the Bogoliubov ansatz, with real $u_i(s)$ and $v_i(s)$,
\begin{eqnarray}
c_i &=& u_i(s)a_i + v_i(s)a^\dagger_i,\nonumber\\
a_i &=& u_i(s)b_i - v_i(s)c^\dagger_i. 
\label{bog}
\end{eqnarray}
The $c_i$-zero occupation state is denoted by $|0_c\rangle$,
where the $s$-dependence of the state is implicit.
The canonicity $[c_i,c^\dagger_j]=\delta_{ij}$ demands
\begin{eqnarray}
u^2_i(s) - v^2_i(s) &=& 1,
\label{canon2}
\end{eqnarray}
upon which
\begin{eqnarray}
\langle a^\dagger_i a_j\rangle &=& v^2_i(s)\delta_{ij},
\nonumber\\
\langle a^\dagger_i a^\dagger_j\rangle &=& \langle a_i a_j
\rangle = -u_i(s) v_i(s)\delta_{ij}, 
\label{exp}
\end{eqnarray}
where
\begin{eqnarray}
\langle A_{k_i}A_{k_j}\rangle &\equiv& \langle 0_c|A_{k_i}A_{k_j}
|0_c\rangle, 
\label{ev}
\end{eqnarray}
We pay particular attention to the following terms
\begin{eqnarray}
:a^\dagger_i a_i:_c &=& \left(u^2_i(s)+v^2_i(s)\right)
c^\dagger_i c_i -
u_i(s) v_i(s)\left(c^{\dagger 2}_i +c^2_i\right),\nonumber\\
:a_i a _i:_c &=& -2u_i(s)v_i(s) c^\dagger_i c_i + v^2_i(s)
 c^{\dagger 2}_i + u^2_i(s) c^2_i. 
\label{sub}
\end{eqnarray}
Needed next is a version of the Wick theorem~\cite{Itzyk} 
for the ordinary product involving the operators $A_1$, ..., $A_n$,
\begin{eqnarray}
A_1\cdots A_n &=& \sum_{p=0}^{[n/2]}:A_1\cdots \hat A_{k_1} \cdots 
\hat A_{k_{2p}}\cdots A_n:_c\nonumber\\
&&\times
\left\{\langle A_{k_1}A_{k_2}\rangle \cdots \langle A_{k_{2p-1}}
A_{k_{2p}}\rangle+ {\rm permutations}\right\}, 
\label{wick}
\end{eqnarray}
where the normal ordering $:\cdots:_c$ is done with respect to some 
annihilation operators $c_i$ which annihilate some common state 
$|0_c\rangle$ for various
$i$, $c_i|0_b\rangle = 0$.  The equation~(\ref{wick}), in which the hatted 
operators are omitted from the normal ordering, is an exact result and 
can be proved by induction.

We can now list the various steps of our estimation:
\begin{enumerate}
\item  We apply the Wick theorem, with respect to $c_i$'s, to the 
hamiltonian ${\cal H}(s)$~(\ref{hamiltonian})
\begin{eqnarray}
{\cal H}(s) &=& E_b(s) I \nonumber\\
&& + \sum_{i}\mbox{\rm terms linear in $c_i$ and $c_i^\dagger$}
\nonumber\\
&& + \sum_i \left(G_i(s)c_i^\dagger c_i + K_i(s) c_i^2 + 
K^*_i(s)c_i^{\dagger 2}\right)\nonumber\\
&& + \sum_{i\not=j} \mbox{\rm terms involving $c_i^\dagger c_j$,
 $c_i c_j$,  $c_i^\dagger c_j^\dagger$}\nonumber\\
&& + \sum_{ijk\cdots}\mbox{\rm higher-order, normal-ordered terms of $c$ 
and $c^\dagger$.} 
\label{final}
\end{eqnarray}
Note that in this process the
higher-order terms contribute to the coefficients of lower-order ones 
through products of various expectation values in~(\ref{exp}). 
With help from~(\ref{exp}) and~(\ref{sub}), the various coefficients 
$E_b(s)$, $G_i(s)$, $K_i(s)$, ...
in the right hand side above can be expressed as polynomials in 
$(u_i(s), v_i(s))$.  
\item  We next fix $u_i(s)$, and $v_i(s)$ numerically from~(\ref{canon2}) 
and from
the imposition that the coefficients $K_i(s)=0$ in~(\ref{final}).
Then we can evaluate the coefficients $G_i(s)$ from their polynomial 
expressions in $(u_i(s),v_i(s))$.  From~(\ref{final}), on the other 
hand, we see also that
\begin{eqnarray}
G_i(s) &=& \langle 1_{c_i}| {\cal H}(s) |1_{c_i}\rangle -
\langle 0_c| {\cal H}(s) |0_c\rangle, 
\label{gap2}
\end{eqnarray}
where $|1_{c_i}\rangle = c^\dagger_i |0_c\rangle$.  Mathematically, 
${\min}_{i}|G_i(s)|$ thus provides some indication of the
size of the energy gaps of ${\cal H}(s)$ around the energy 
level $E_b(s) = \langle 0_c| {\cal H}(s) |0_c\rangle.$  
Even though $|0_c\rangle$, obtained from the linear Bogoliubov
transformations~(\ref{bog}), in general may not be the true ground 
state of ${\cal H}(s)$, 
we could use ${\min}_{i}|G_i(s)|$ as some indicator for the gap 
$g$ in~(\ref{gap}) in order to estimate $T$ from the
adiabaticity condition~(\ref{adiab}).
\end{enumerate}

Some variations of the above method might yield a better estimate.  For
instance, one could numerically obtain $(u_i(s), v_i(s))$, and thus 
${\min}_{i}|G_i(s)|$, by
minimising the energy $E_b(s)$ subjected to the constraints~(\ref{canon2})
with or without the constraints $K_i(s) = 0$.
% that is we could minimise
%\[E_b(s) +\lambda\sum_i\left((u_i^2-v_i^2 -1)^2 + K_i^2(s) \right)\]
%Swith respect to $(u_i, v_i)$, where $\lambda$ is the Lagrange multiplier.
The aim of this minimisation, if possible, is to select a state $|0_c\rangle$
whose energy expectation value at the time $s$ is as closed to that of the
ground state at that instant as allowed by the linear 
Bogoliubov transformations.  Other variations might be involving,
instead of~(\ref{bog}), some non-linear relations, for the
canonicity only requires that $a$ and $c$ are to be unitarily transformed,
$c = U^\dagger(a^\dagger,a)aU(a^\dagger,a)$.

The accuracy of the estimation and its higher order corrections 
can be evaluated systematically from the higher order terms in the 
last line of~(\ref{final}). 
Numerical diagonalisation of~(\ref{final}) 
with a series of truncated bases in $|n_{c_i}\rangle$ of increasing 
sizes could further provide us more information about the gap thus obtained.

% It turns out that with the method above
% we have traded the original Diophantine equation for a set
% of polynomials equations in $(u_i(s),v_i(s))$.  
% The solutions of these latter equations, however, are real,
% not integers as in the case of the original Diophantine, and could be 
% obtained or well approximated through the calculus of continuous variables.  
% (Tarski~\cite{Tarski} has found that looking for the existence
% of real solutions of polynomials over the real numbers are less 
% restrictive than that of integer solutions of Diophantine equations; 
% he has shown that arithmetics over the reals is, in fact, decidable.)  
% Note that in some cases we may not even need to solve the equations exactly 
% in order to have some appropriate estimate for ${\min}_{i}|G_i(s)|$.

% More general, we have traded the hard question of integer solutions for
% some more palatable estimation for the real-valued
% time $T$ in~(\ref{adiab}).  This estimation is less restrictive in the
% sense that it needs only be of order of magnitude for a good success
% probability, which could be boosted further.

\end{document}